\documentclass[conference]{IEEEtran}
\usepackage{cite}
\usepackage{lscape}
\usepackage{pdfpages}
\usepackage[figuresleft]{rotating}
\usepackage{amsmath,amssymb,amsfonts}
\usepackage{algorithmic}
\usepackage{graphicx}
\usepackage{subcaption}
\usepackage{textcomp}
\usepackage{xcolor}
\usepackage{soul}
\usepackage{multirow}
\usepackage{booktabs}
\usepackage{adjustbox}
\usepackage{mathtools}
\def\BibTeX{{\rm B\kern-.05em{\sc i\kern-.025em b}\kern-.08em
    T\kern-.1667em\lower.7ex\hbox{E}\kern-.125emX}}

\usepackage{tikz}
\usepackage{lipsum}

\newcommand\copyrighttext{%
    \centering
  \tiny
  \copyright~2020 IEEE. Personal use of this material is permitted.  Permission from IEEE must be obtained for all other uses, in any current or future media, including reprinting/republishing this material for advertising or promotional purposes, creating new collective works, for resale or redistribution to servers or lists, or reuse of any copyrighted component of this work in other works.}
\newcommand\copyrightnotice{%
\begin{tikzpicture}[remember picture,overlay]
\node[anchor=south,yshift=10pt] at (current page.south) {\fbox{\parbox{\dimexpr\textwidth-\fboxsep-\fboxrule\relax}{\copyrighttext}}};
\end{tikzpicture}%
}

\begin{document}

\title{Acoustic Scene Classification with Squeeze-Excitation Residual Networks}

\author{\IEEEauthorblockN{Javier Naranjo-Alcazar}
\IEEEauthorblockA{
\textit{Visualfy, Universitat de Valencia}\\
Benisano, Spain \\
javier.naranjo@visualfy.com}
\and
\IEEEauthorblockN{Sergi Perez-Castanos}
\IEEEauthorblockA{
\textit{Visualfy}\\
Benisano, Spain \\
sergi.perez@visualfy.com}
\and
\IEEEauthorblockN{Pedro Zuccarello}
\IEEEauthorblockA{
\textit{Visualfy}\\
Benisano, Spain \\
pedro.zuccarello@visualfy.com}
\and
\IEEEauthorblockN{Maximo Cobos}
\IEEEauthorblockA{
\textit{Universitat de Valencia}\\
Burjassot, Spain \\
maximo.cobos@uv.es}
}

\maketitle

\begin{abstract}
Acoustic scene classification (ASC) is a problem related to the field of machine listening whose objective is to classify/tag an audio clip in a predefined label describing a scene location (e. g. park, airport, etc.).
Many state-of-the-art solutions to ASC incorporate data augmentation techniques and model ensembles. However, considerable improvements can also be achieved only by modifying the architecture of convolutional neural networks (CNNs). In this work we propose two novel squeeze-excitation blocks to improve the accuracy of a CNN-based ASC framework based on residual learning. The main idea of squeeze-excitation blocks is to learn spatial and channel-wise feature maps independently instead of jointly as standard CNNs do. This is usually achieved by some global grouping operators, linear operators and a final calibration between the input of the block and its obtained relationships. The behavior of the block that implements such operators and, therefore, the entire neural network, can be modified depending on the input to the block, the established residual configurations and the selected non-linear activations. The analysis has been carried out using the TAU Urban Acoustic Scenes 2019 dataset\footnote{https://zenodo.org/record/2589280} presented in the 2019 edition of the DCASE challenge.  All configurations discussed in this document exceed the performance of the baseline proposed by the DCASE organization by 13\% percentage points. In turn, the novel configurations proposed in this paper outperform the residual configurations proposed in previous works.
\end{abstract}


\copyrightnotice

\section{Introduction}\label{sec:intro}

\par The analysis of everyday ambient sounds can be very useful when developing intelligent systems in applications such as domestic assistants, surveillance systems or autonomous driving. Acoustic scene classification (ASC) is one of the most typical problems related to machine listening \cite{valenti2017convolutional, bae2016acoustic, pham2020robust, han2016acoustic}. Machine listening is understood as the field of artificial intelligence that attempts to create intelligent algorithms capable of extracting meaningful information from audio data. Therefore, ASC can be defined as the area of machine listening that attempts to tag an audio clip in one of the predefined tags related to the description of a scene (for example, airport, park, subway, etc.). 




\par The first approaches to the ASC problem were centered on the design of proper inputs to the classifier, this is, feature engineering \cite{martin2016case}. Most research efforts tried to create meaningful representations of the audio data to later feed gaussian mixture models (GMMs), hidden Markov models (HMM) or support vector machines (SVMs) \cite{MartinTASL}. In this context, a wide range of input representations were proposed such as Mel-frequency cepstral coefficients (MFCCs) \cite{Roma2013, Li2013}, Wavelets \cite{Li2013}, constant-Q transform (CQT) or histograms of oriented gradientes (HOG) \cite{Rakotomamonjy2013}, among others. 

\par With the years and the emergence of convolutional networks in the field of image, the most common option among machine listening investigations was the implementation of a CNN generally fed with a 2D audio representation, usually a log-Mel spectrogram \cite{valenti2017convolutional, mesaros2019acoustic}. These networks have shown very satisfactory results, especially when they are trained with large datasets. This is why data augmentation techniques are commonly applied, such as mixup strategies \cite{zhang2017mixup} or temporal cropping \cite{Gao2019}. In addition, to improve the final accuracy, many studies use ensembles, combining the output from different classifiers to obtain a single more robust prediction. Unfortunately, the use of ensembles makes it more difficult to analyze the contribution to the classification performance of a new CNN architecture integrated within the proposed ensemble. To avoid such issue, this work considers isolated contributions of several CNN architectures implemented with different residual blocks based on squeeze-excitation methods, without any extra modifications during the training or inference phases.

\par CNNs are built with stacked convolutional layers. These layers learn its filter coefficients by capturing local spatial relationships (neighbourhood information) along the input channels and generate features maps (filtered inputs) by jointly encoding the spatial and channel information. In all application domains (image classification/segmentation, audio classification/tagging, etc.), the idea of encoding the spatial and the channel information independently is less studied, although it has shown promising results \cite{roy2018concurrent, hu2018squeeze}.



\par In order to provide insight about the behaviour of CNNs when analyzing spatial and channel information independently, several squeeze-excitation (SE) blocks have been presented in the image classification literature \cite{hu2018squeeze,roy2018concurrent}. In \cite{hu2018squeeze}, a block that ``squeezes" spatially and ``excites" channel-wise with linear relationships was presented. The idea behind this block, denoted as \emph{cSE} in this work, is to model the interdependencies between the channels of feature maps by exciting in a channel-wise manner. This type of block showed its effectiveness in image classification tasks, surpassing state-of-the-art networks only by inserting it in a specific point of the network. Following this idea, two more blocks were presented in \cite{roy2018concurrent}. The first one, denoted as \emph{sSE}, ``squeezes" along the channels and ``excites" spatially, whereas the last block, \emph{scSE}, combines both strategies. The scSE block recalibrates the feature maps along spatial and channel dimensions independently (cSE and sSE) and then combines the information of both paths by adding their outputs. This last block showed the most promising results in image domain tasks. According to \cite{roy2018concurrent}, this block forces the feature maps to be more informative, both spatially and channel-wise.

In this work we propose two novel residual squeeze-excitation blocks plus an analysis of different state-of-the art configurations for addressing the ASC problem. These two new configurations are intended to enhance the benefits of residual training and recalibration of feature maps using squeeze-excitation techniques jointly. Squeeze-excitation techniques allows the network to extract more meangniful information during training and residual learning helps the training procedure to run it in a more efficient manner. The results show that, by using the proposed novel configurations, results are considerably improved. Moreover, it is shown that all the residual squeeze-excitation configurations perform better than a classical convolutional residual block.

\par The following of the paper is organized as follows. Sect.~\ref{sec:background} provides further insight into the techniques to be analyzed. Sect.~\ref{sec:res_blocks} introduces the different squeeze-excitation blocks proposed in this work. Section \ref{sec:exp_details} describes the dataset used for validating the premises previously explained plus the audio pre-processing in order to feed the CNN. Section \ref{sec:results} shows the experimental results and Section \ref{sec:conclusion} concludes our work.



\section{Background}
\label{sec:background}

\subsection{Squeeze-Excitation blocks}\label{sec:se_blocks}

\par Squeeze-excitation (SE) blocks can be understood as modules for channel recalibration of feature maps \cite{roy2018concurrent}. Let' s assume an input feature map, $\mathbf{X} \in \mathbb{R}^{H \times W \times C'}$, that feeds any convolutional block, usually implemented by convolutional layers and non-linearities, and generates an output feature map $\mathbf{U} \in \mathbb{R}^{H \times W \times C}$. Here, $\mathbf{U}$ could also be expressed as $\mathbf{U} = [\mathbf{u}_{1}, \mathbf{u}_{2}, \dots, \mathbf{u}_{C}]$, being $\mathbf{u}_{i} \in \mathbb{R}^{H \times H}$ a channel output. Considering this notation, $H$ and $W$ represents the height and the width, while $C'$ and $C$ defines the number of input and output channels, respectively. The convolutional process function can be defined as $\mathbf{F(\cdot)}$, so that $\mathbf{F}(\mathbf{X}) = \mathbf{U}$. The output $\mathbf{U}$ is generated by combining the spatial and channel information of $\mathbf{X}$. The objective of SE blocks is to recalibrate $\mathbf{U}$ with $\mathbf{F}_{SE}(\cdot)$ to generate $\hat{\mathbf{U}}$, i.e.  $\mathbf{F}_{SE}(\cdot): \mathbf{U} \rightarrow \hat{\textbf{U}}$. This recalibrated feature map, $\hat{\mathbf{U}}$, can be stacked after every convolutional block and then used as input to the forthcoming pooling layers. This recalibration can be carried out wit different types of block functions $F_{SE}(\cdot)$, as explained in the following subsections.

\subsubsection{Spatial Squeeze and Channel Excitation Block (cSE)}\label{subsubsec:cse}

In a cSE module (depicted in Fig.~\ref{fig:SEblocks}(a)) for spatial squeeze and channel excitation, a unique feature map of each channel from $\mathbf{U}$ is ﬁrst obtained by means of global average pooling. This operator produces a vector $\mathbf{z} \in \mathbb{R}^{1 \times 1 \times C}$. The $k$th element of such vector can be expressed as:

\begin{equation}
z_{k} = \frac{1}{H \times W} \sum_{i}^{H} \sum_{j}^{W} \mathbf{u}_{k}(i, j), \quad k = 1,\dots,C,
\label{eq:global}
\end{equation}
where $\mathbf{u}_k(i,j)$ denotes the $(i,j)$ element of the $k$th channel feature map.

As suggested by Eq.~(\ref{eq:global}), global spatial information is embedded in vector $\mathbf{z}$. This representation is then used to extract channel-wise dependencies using two fully-connected layers, obtaining the transformed vector $\hat{\mathbf{z}}$. Therefore, $\hat{\mathbf{z}}$ can be expressed as $\hat{\mathbf{z}} = \mathbf{W}_{1}(\delta(\mathbf{W}_{2} \mathbf{z}))$, where $\delta$ represents ReLU activation. $\mathbf{W}_{1} \in \mathbb{R}^{C \times \frac{C}{\rho}}$ and $\mathbf{W}_{2} \in \mathbb{R}^{\frac{C}{\rho} \times C}$ are the weights of the fully-connected layers, and $\rho$ is a ratio parameter. As last step, the activation range is compressed to the interval $[0, 1]$ using a sigmoid activation function, $\sigma$. This final step indicates the importance of each channel and how they should be rescaled. The purpose of this recalibration is to let the network ignore channels with less information and emphasize the ones that provide more meaningful information. Then, the rescaled feature maps, $\hat{\mathbf{U}}$, can be expressed as \cite{hu2018squeeze, roy2018concurrent}:

\begin{equation}\label{eq:cse}
    \hat{\mathbf{U}}_{cSE} = F_{cSE}(\mathbf{U}) = [\sigma (\hat{z}_{1})\mathbf{u}_{1}, \dots, \sigma (\hat{z}_{C})\mathbf{u}_{C}],
\end{equation}
 where $\hat{z}_k$ are the elements of the transformed vector $\hat{\mathbf{z}}$.
 
\subsubsection{Channel Squeeze and Spatial Excitation Block (sSE)}\label{subsubsec:sse}

In the case of an sSE block \cite{roy2018concurrent}, as shown in Fig.~\ref{fig:SEblocks}(b), a unique convolutional layer with one ﬁlter and $(1,1)$ kernel size is implemented to obtain a channel squeeze and spatial excitation effect. Here, it is assumed an alternative representation of the input tensor as $\mathbf{U} = [\mathbf{u}^{1,1}, \mathbf{u}^{1,2}, \dots ,\mathbf{u}^{i,j}, \dots, \mathbf{u}^{H,W}]$ where $\mathbf{u}^{i,j} \in \mathbb{R}^{1 \times 1 \times C}$. The convolution can be expressed as $\mathbf{q} = \mathbf{W} \star \mathbf{U}$, being $\textbf{W} \in \mathbb{R}^{1 \times 1 \times C \times 1}$ and $\textbf{q} \in \mathbb{R}^{H \times W}$. Each $q_{i, j}$ represents the combination of all channels in location $(i, j)$. As done with cSE, the output of this convolution is passed through a sigmoid function. Each $\sigma(q_{i, j})$ determines the importance of the specific location $(i, j)$ across the feature map. Like the previous block, this recalibration process indicates which locations are more meaningful during the training procedure. As a result, the output of the SE block can be expressed as \cite{roy2018concurrent}:

\begin{equation}\label{eq:sse}
    \hat{\mathbf{U}}_{sSE} = F_{sSE}(\textbf{U})= [\sigma(q_{1, 1})\textbf{u}^{1,1}, \dots, \sigma(q_{H, W})\textbf{u}^{H,W}].
\end{equation}

\subsubsection{Spatial and Channel Squeeze \& Excitation block (scSE)}\label{subsubsec:scSE}

\begin{figure}[!t]
\centering
\includegraphics[width=0.8\columnwidth]{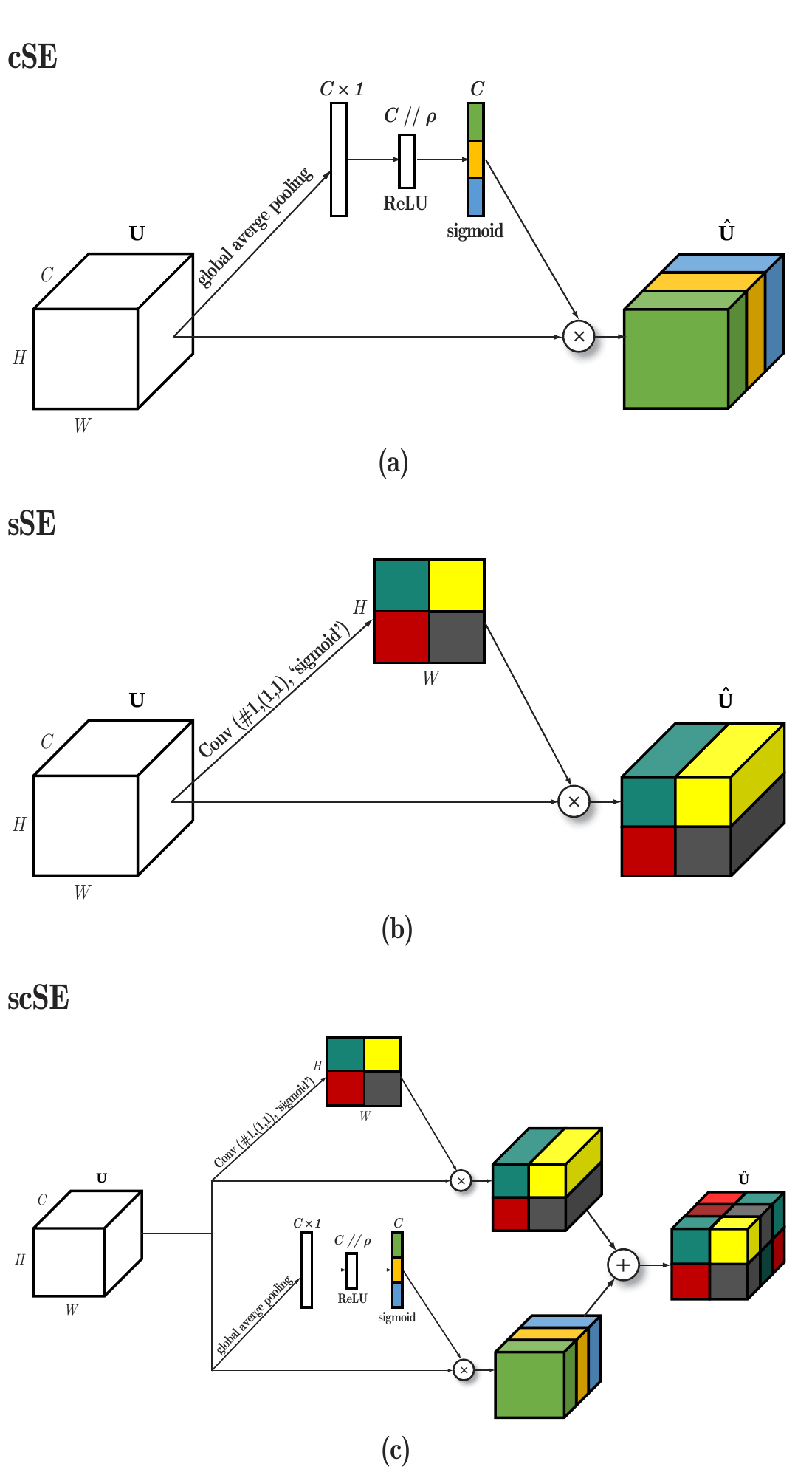}
\caption{Diagram of different SE blocks: (a) describes cSE block procedure, (b) ilustrates sSE block framework and (c) shows scSE block by combining (a) and (b).}
\label{fig:SEblocks}
\end{figure}

The scSE block \cite{roy2018concurrent} is implemented by declaring cSE and sSE blocks in parallel and adding both outputs (see Fig.~\ref{fig:SEblocks}(c)). It has been reported that the scSE block shows better performance than cSE and sSE used independently. In this case, a location $(i, j, c)$ gets a higher sigmoid or activation value when both channel and spatial recalibration get it at the same time. In this case, the network focuses on feature maps that are meaningful from both a spatial and channel-wise point of view. Formally, this SE block can be defined as \cite{roy2018concurrent}

\begin{equation}\label{eq:cssse}
    \hat{\mathbf{U}}_{scSE} = \hat{\mathbf{U}}_{cSE} + \hat{\mathbf{U}}_{sSE}.
\end{equation}

\subsection{Residual Networks}\label{subsec:res_net}

Residual networks were first proposed in \cite{he2016deep}. A network of this kind replaces the standard stacked convolutional layers \cite{simonyan2014very} by residual blocks. Residual layers are designed to approximate a residual function: $\mathcal{F}(\mathbf{X}) \coloneqq \mathcal{H}(\mathbf{X})-\mathbf{X}$, where $\mathcal{H}$ represents the mapping to be fit by a set of stacked layers and $\mathbf{X}$ represents the input to the first of such stacked layers. The original function $\mathcal{H}$ can therefore be defined as $\mathcal{H}(\mathbf{X})~=~\mathcal{F}(\mathbf{X}) + \mathbf{X}$. The main motivation of choosing this kind of network corresponds to the intuition that optimizing a residual mapping may be easier than optimizng the original unreferenced one, as in a classical convolutional network. A simple way of implementing residual learning in CNNs is by adding a shortcut connection that performs as an identity mapping, adding back the input $\mathbf{X}$ to the output of the residual block $\mathcal{F}(\mathbf{X})$. In the first proposition of the residual block, Rectified ReLU activation is applied after the addition and the result of such activation becomes the input for the next residual block. Note, that in the first configuration, shortcut connections do not add more parameters nor extra computational cost. Therefore, deeper networks can be trained with little additional effort, reducing vanishing-gradient problems. As it will be later explained, in this work, the identity mapping is replaced with a $1 \times 1$ convolutional layer as it is explained in Section \ref{sec:res_blocks}. Therefore, this work function can be expressed as $\mathcal{H}(\mathbf{X})~=~\mathcal{F}(\mathbf{X}) + g(\mathbf{X})$ where $g()$ represents the convolutional process with the learnt filter coefficients.

\section{Configurations for Squeeze-and-Excitation Residual Networks}\label{sec:res_blocks}

According to \cite{hu2018squeeze}, SE blocks exhibit better performance when deployed on networks with residual configuration than on VGG-style networks. Therefore, two novel residual blocks implementing scSE modules are presented in this paper. The performance of these two newly proposed blocks is compared against other state-of-the-art residual configurations that incorporate SE modules.

\subsection{SE Block Description}

\begin{figure*}[h]
    \centering
    \centerline{\includegraphics[width=0.9\textwidth]{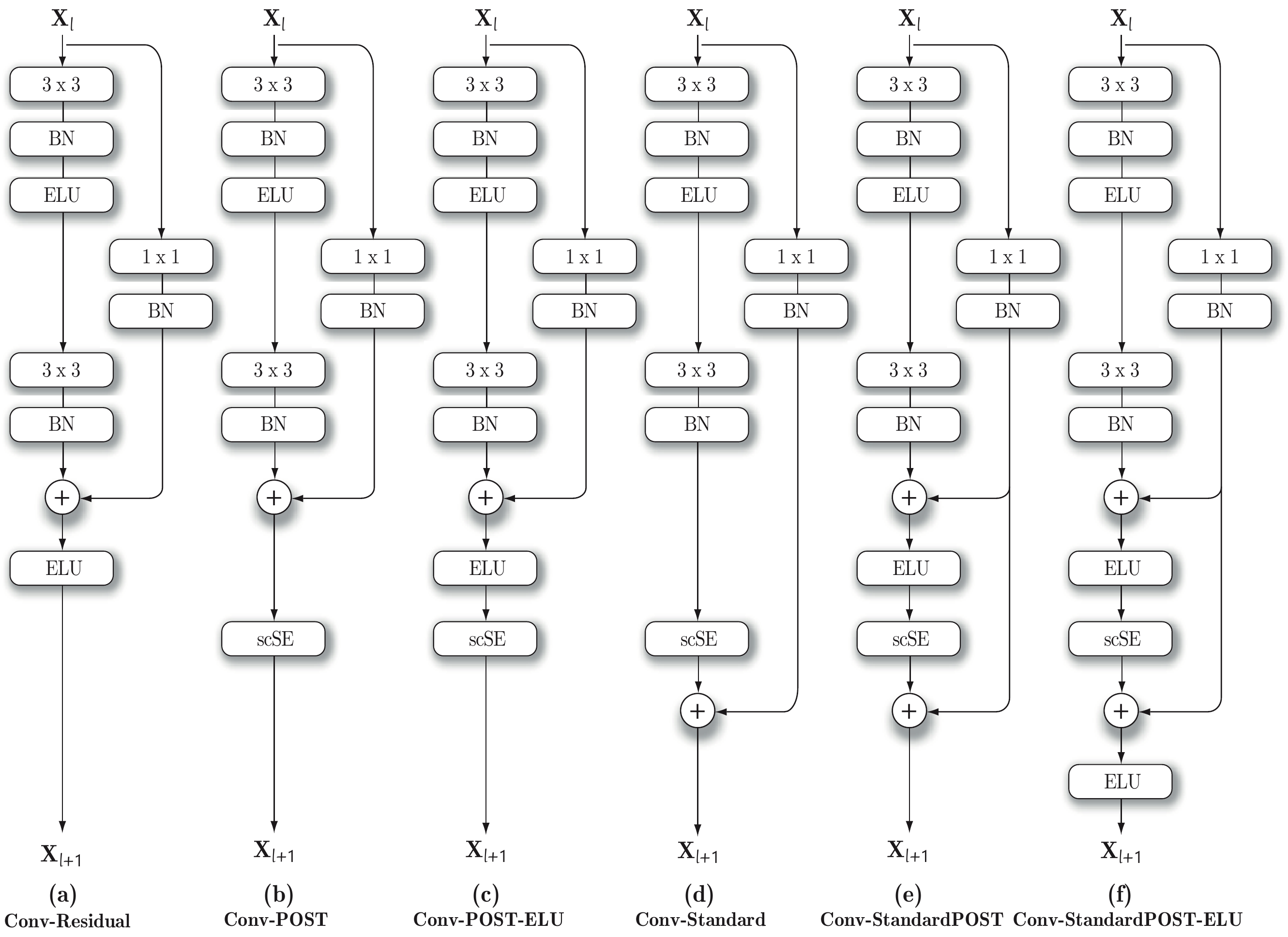}}
    \caption{Different residual squeeze-excitation blocks analyzed in this work: (a) is inspired by the first residual block proposed in \cite{he2016deep}; (b), (c) and (d) are inspired by the work done in \cite{hu2018squeeze}; (e) and (f) are the two novel configurations proposed in this work.}
    \label{fig:squeeze_2d}
\end{figure*}

All the configurations analyzed in this work are depicted in Figure \ref{fig:squeeze_2d}. In the following, we describe in details these blocks.

\paragraph{Conv-residual} shown in Fig.~\ref{fig:squeeze_2d}(a), is inspired by \cite{he2016deep}. It is used as a baseline in order to validate the network performance without any squeeze-excitation and how much it can be improved when incorporating these blocks. In the present work some slight modifications for a more convenient implementation were introduced: the shortcut connection was implemented with a $1\times 1$ convolutional layer and the activation after the addition was set to an exponential linear unit (ELU) function \cite{shah2016deep, clevert2015fast}.

\paragraph{Conv-POST} shown in Fig.~\ref{fig:squeeze_2d}(b), is inspired by the block referred to as \emph{se-POST} in \cite{hu2018squeeze}. The scSE block is included at the end and is equivalent to a recalibration of the Conv-residual block.

\paragraph{Conv-POST-ELU} shown in Fig.~\ref{fig:squeeze_2d}(c), is very similar to the above Conv-POST block, but the recalibration is performed over the ELU-activated output of the residual block.

\paragraph{Conv-Standard} shown in Fig.~\ref{fig:squeeze_2d}(d), is inspired by \cite{hu2018squeeze}, where the scSE block is stacked after the convolutional block for recalibrating prior to adding the shortcut branch.

\paragraph{Conv-StandardPOST} shown in Fig.~\ref{fig:squeeze_2d}(e) is proposed in this work to create a double shortcut connection, one before SE calibration and one after. The idea is to let the network learn residual mappings simultaneously with and without SE recalibration, thus, affecting the way in which the block optimizes the residual by considering jointly standard and post SE-calibrated outputs.

\paragraph{Conv-StandardPOST-ELU} shown in Fig.~\ref{fig:squeeze_2d}(f) is the other proposed block, corresponding to the above explained Conv-StandardPOST block, but followed by ELU activation.

To summarize, the output $\mathbf{X}_{l+1}$ of each block for an input $\mathbf{X}$ is given by:
\begin{eqnarray}
\mathrm{a)} \quad \mathbf{X}_{l+1} &=& \mathcal{R}\left( \mathcal{H}(\mathbf{X}) \right), \\
\mathrm{b)} \quad \mathbf{X}_{l+1} &=& \mathbf{F}_{SE}\left(\mathcal{H}(\mathbf{X}) \right), \\
\mathrm{c)} \quad \mathbf{X}_{l+1} &=& \mathbf{F}_{SE}\left(\mathcal{R}\left(\mathcal{H}(\mathbf{X})\right) \right),\\
\mathrm{d)} \quad \mathbf{X}_{l+1} &=& \mathbf{F}_{SE}\left(\mathcal{F}(\mathbf{X}) \right) + g(\mathbf{X}),\\
\mathrm{e)} \quad \mathbf{X}_{l+1} &=& \mathbf{F}_{SE}\left(\mathcal{H}(\mathbf{X})\right) + g(\mathbf{X}),\\
\mathrm{f)} \quad \mathbf{X}_{l+1} &=& \mathcal{R}\left( \mathbf{F}_{SE}\left(\mathcal{H}(\mathbf{X})\right) + g(\mathbf{X})\right),
\end{eqnarray}
where $\mathcal{R}(\cdot)$ refers to ELU activation function with $\alpha$ parameter set to 1. As it will be discussed in \ref{sec:results}, the two proposed configurations have shown to outperform the rest in the considered acoustic scene analysis task.

\par In order to avoid possible duplications or expansion processes in the channel dimension, the identity branch is replaced by a convolutional layer with a $(1,1)$ kernel size and with the same number of filters as the residual branch. Including such convolutional layer in the shortcut branch creates a projection that avoids dimensionality conflicts in the residual block addition.

\subsection{Network Architecture}\label{sec:metod}

\par The CNN implemented in order to validate the behaviour of the different squeeze-excitation configurations has been inspired on \cite{perezcastanos2019cnn} where a VGG-style \cite{simonyan2014very} network with 3 convolutional blocks followed by different max-pooling and dropout \cite{srivastava2014dropout} operators is implemented. In the present work, the original convolutional blocks have been replaced with the different residual squeeze-excitation blocks proposed in this study. The max-pooling, dropouts and linear layers are configured with the same parameters as in \cite{perezcastanos2019cnn}. The network architecture can be found in Table \ref{tab:net_arch}.


\begin{table}[]
\caption{Proposed network for validating the scSE configurations of Fig.~\ref{fig:squeeze_2d}. Values preceeded by \# correspond to the number of filters. Kernel sizes are set as indicated in Fig.~\ref{fig:squeeze_2d}. This architecture is inspired by the work in \cite{perezcastanos2019cnn}}
\centering
\begin{tabular}{c}
\toprule
Residual-scSE block (\#32, $\rho$ = 2)  \\
\midrule
MaxPooling(2,10)                                        \\
\midrule
Dropout(0.3)                                            \\
\midrule
Residual-scSE block (\#64, $\rho$ = 2) \\
\midrule
MaxPooling(2,5)                                         \\
\midrule
Dropout(0.3)                                            \\
\midrule
Residual-scSE block (\#128, $\rho$ = 2)    \\
\midrule
MaxPooling(2,5)                                         \\
\midrule
Dropout(0.3)                                            \\
\midrule
Flatten                                                 \\
\midrule
{[}Dense(100), batch normalization, ELU{]}          \\
\midrule
Dropout(0.4)                                            \\
\midrule
{[}Dense(10), batch normalization, softmax{]}  \\
\bottomrule
\end{tabular}
\label{tab:net_arch}
\end{table}

\par As the database used in the current work is much smaller than the one in \cite{hu2018squeeze}, some of the hyperparameters that define the components of the scSE block had to be modified. The number of elements in the Dense layer with ReLU activation in Fig.~\ref{fig:SEblocks}(a) has been set to 16 in the first Residual-scSE block, the same as in \cite{hu2018squeeze} in its cSE block, but the number of filters at the input, $C$, has been set to $C=32$. Therefore, the ratio between these parameters throughout the network is two, as observed Table \ref{tab:net_arch}.

\begin{figure*}[h]
    \centering
    \centerline{\includegraphics[scale=0.17]{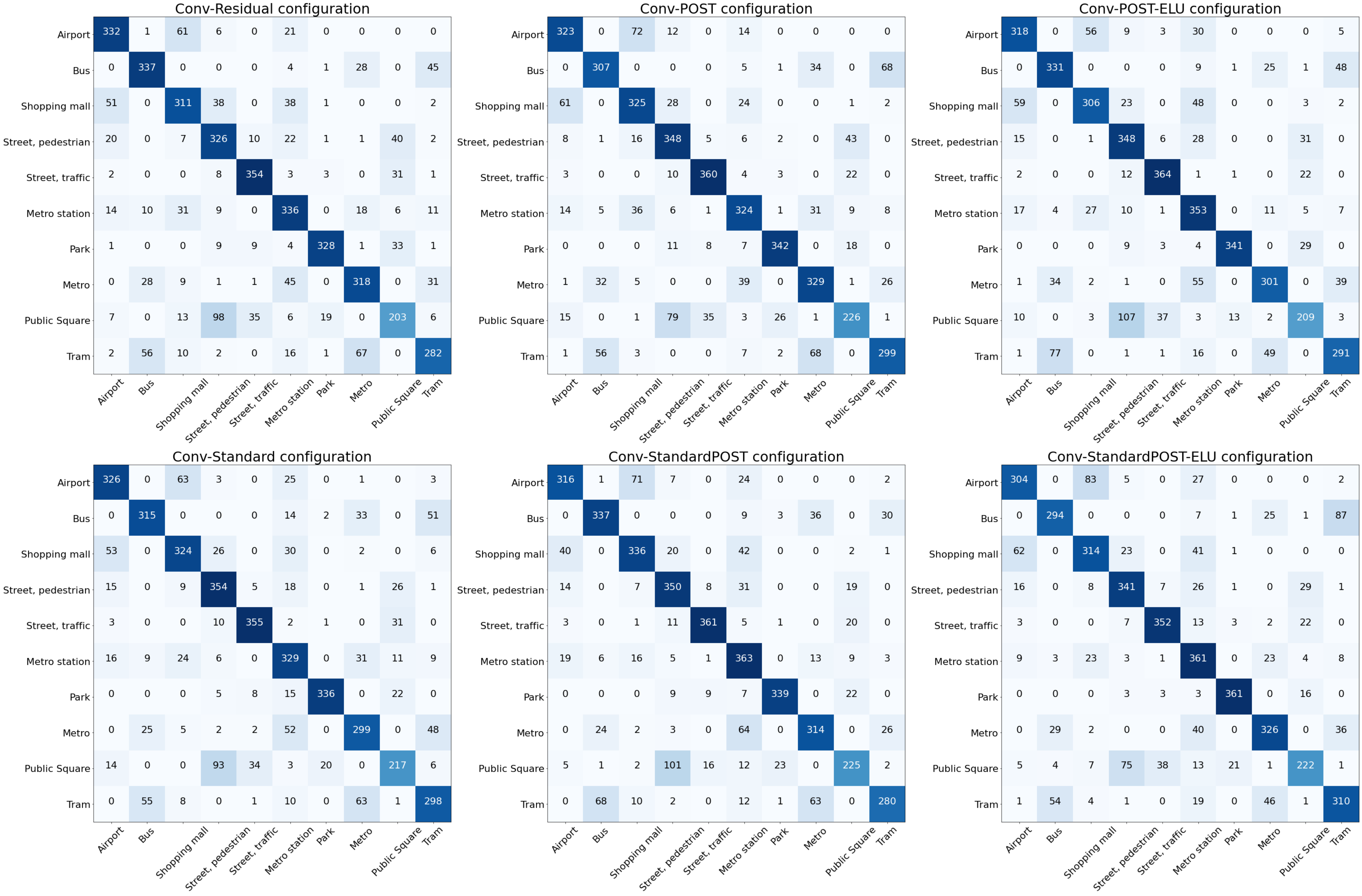}}
    \caption{Confusion matrices for the generated models over the evaluation dataset.}
    \label{fig:conf_matrices}
\end{figure*}

\section{Experimental details}\label{sec:exp_details}

This section describes in detail the experimental implementation carried out to conduct the analysis of the presented SE residual blocks, including the datasets, the audio representation selected to feed the network and the training configuration.

\subsection{Dataset}\label{subsec:dataset}

\par To check the behavior of these implementations in an ASC problem, the TAU Urban Acoustic Scenes 2019, Development dataset presented in Task 1A of the 2019 edition of DCASE has been used \cite{mesaros2019acoustic}. The database consists of 40 hours of stereo audio-recording in different urban environments and landscapes such as parks, metro stations, airports, etc. making a total of 10 different scenes. These have been recorded in different cities such as Barcelona, Paris or Helsinki, among others. All  audio clips are 10-second long. They are divided into two subsets of 9185 and 4185 clips for training and validation, respectively.

\subsection{Audio processing}

\par The input to the network is a 2D log-Mel spectrogram representation with 3 audio channels. The three channels are composed of the harmonic and percussive component \cite{fitzgerald2010harmonic, driedger2014extending} of the signal converted to mono and the difference between left ($L$) and right ($R$) channels. That is, the first channel corresponds to the log-Mel spectrogram of the harmonic source, the second channel corresponds to the same representation but over the percussive source and the last one to the log-Mel spectrogram of the difference between channels calculated by subtracting left and right channels $(L-R)$. This representation, known as HPD, was presented in \cite{perezcastanos2019cnn}. The log-Mel spectrogram is calculated using 64 Mel filters with a 
window size of 40~ms and 50\% overlap. Therefore, an audio clip becomes a $64\times T \times3$ array with $T$ being the number of time frames. In this specific dataset, the input audio representation corresponds to an array of dimension $64\times500\times3$.

\subsection{Training procedure}\label{subsec:training_procedure}

The training process was optimized using the Adam optimizer \cite{kingma2014adam}. The cost function used was the categorical crossentropy. Training was limited to a maximum of 500 epochs but early stopping is applied if the validation accuracy does not improve by 50 epochs. If this same metric does not improve in 20 epochs, the learning rate is decreased by a factor of 0.5. The batch size used was 32 samples.


\section{Results}\label{sec:results}

\begin{table}[]
\caption{Accuracy results from the validation partition in development phase.}
\centering
\begin{tabular}{cc}
\toprule
System                & Development accuracy (\%)    \\
\midrule \midrule
Baseline \cite{mesaros2019acoustic}             & 62.5                    \\
\midrule
Wang\_NWPU\_task1a \cite{wan2019ciaic}    & 72.4 \\
\midrule
Fmta91\_KNToosi\_task1a \cite{arabnezhadurban} & 70.49  \\
\midrule
MaLiu\_BIT\_task1a  \cite{maacoustic}  & 76.1 (evaluation)       \\
\midrule
DSPLAB\_TJU\_task1a \cite{Ding2019}  & 64.3                    \\
\midrule
Kong\_SURREY\_task1a \cite{Kong2019} & 69.2                    \\
\midrule
Liang\_HUST\_task1a \cite{Liang2019}   & 70.70                   \\
\midrule
Salvati\_DMIF\_task1a \cite{Salvati2019}      & 69.7 \\ 
\midrule
 \textit{Conv-Residual}     & \textit{74.51 $\pm$ 0.65} \\ 
\midrule
\textit{Conv-Standard}     & \textit{75.16 $\pm$ 0.33} \\ 
\midrule
\textit{Conv-POST}      & \textit{75.84$\pm$0.65} \\
\midrule
\textit{Conv-POST-ELU}       & \textit{75.81$\pm$0.47} \\ 
\midrule
\textit{Conv-StandardPOST}      & \textit{\textbf{76.72$\pm$0.59}} \\ 
\midrule
\textit{Conv-StandardPOST-ELU}    & \textit{76.00$\pm$0.55} \\ 
\bottomrule
\end{tabular}
\label{tab:results}
\end{table}

\par In order to analyze the contributions of this work with respect to other state-of-the-art approaches, the results obtained with the different configurations presented in this work (see Fig.~\ref{fig:squeeze_2d}) are compared to the ones obtained by different authors in Task 1A of DCASE 2019 using the same dataset. For a fair comparison, only submissions not making use of data augmentation techniques are considered. In the case of submissions that presented an ensemble of several models, only the results of the best performing model making up the ensemble are taken into account. For example, in \cite{wan2019ciaic} a global development accuracy of 78.3\% is reported, but that value was obtained by averaging 5 models. The best individual model obtained 72.4\%, so this is the value presented in Table \ref{tab:results}. This said, please be aware that the accuracy of the final submission\footnote{http://dcase.community/challenge2019/task-acoustic-scene-classification-results-a} may differ from that presented in Table~\ref{tab:results}. Next, we summarize some important features of the competing approaches.

\begin{itemize}
  \item \textbf{Wang\_NWPU\_task1a} \textbf{\cite{wan2019ciaic}}: the audio representation considers two channels using a log-Mel Spectrogram from harmonic and percussive sources similar to our representation. The number of Mel filters is set to 256. The window size is set to 64 ms and the hop size to 15 ms. Mel filters are calculated with cutoff frequencies from 50 Hz to 14 kHz. A VGG-style CNN \cite{simonyan2014very} is used as a classifier.
  \item \textbf{Fmta91\_KNToosi\_task1a} \textbf{\cite{arabnezhadurban}}: wavelet scattering spectral features are extracted from the mono audio signal. A random subspace method is used as classifier.
  \item \textbf{MaLiu\_BIT\_task1a} \textbf{\cite{maacoustic}}: Deep Scattering Spectra features (DSS) are extracted from each stereo channel. Classification is performed with a Convolutional Recurrent Neural Network (CRNN). For this network, Table~\ref{tab:results} does not report the accuracy on the development set (only on the evaluation set). This is because of some mismatch reported by the authors in the validation procedure with the configuration of the dataset. 
  \item \textbf{DSPLAB\_TJU\_task1a} \textbf{\cite{Ding2019}}: this submission approaches the problem in a more classical way extracting audio statistical features such as ZRC, RMSE, spectrogram centroid, etc. A GMM is used as a classifier.
  \item \textbf{Kong\_SURREY\_task1a} \textbf{\cite{Kong2019}}: this submssion can be defined as the state-of-the-art framework in ASC problem. The audio representation considers also the log-Mel spectrogram. The classifier is a VGG-based \cite{simonyan2014very} CNN. This network is a fully convolutional network with no linear layers implemented. The feature maps are reshaped into a one dimensional vector using a global average pooling before the decision layer.
  \item \textbf{Liang\_HUST\_task1a} \textbf{\cite{Liang2019}}: in this method, the log-Mel spectrogram is first extracted after converting the audio signal to mono. Interestingly, the log-Mel spectrogram is divided into two-seconds spectrograms, that means that spectrogram shapes change from [$F \times T \times 1$] to [$F \times (T/5) \times 1$]. This configuration allows training with audio samples consisting of 5 different spectrograms instead of one. A CNN with frequency attention mechanism is implemented as classifier. For more detail of the attention implementation, see \cite{Liang2019}.
  \item \textbf{Salvati\_DMIF\_task1a} \textbf{\cite{Salvati2019}}: unlike the other submissions, this one works directly on the audio vector. To this end, a 1D convolutional network is implemented. Although some recent efforts have been made in this direction \cite{naranjoalcazar2019performance}, the state-of-the-art literature shows that 2D audio representations, such as spectrograms, still obtain the better classification results \cite{dieleman2014end}.
  \item \textbf{DCASE baseline} \textbf{\cite{mesaros2019acoustic}}: the audio is first converted to mono and a log-Mel spectrogram is extracted. In this case, only 40 Mel bins are calculated instead of 64, which is the typical state-of-the-art choice. A CNN is used as a classifier with 2 convolutional layers. The 1D conversion before classification layers is performed by a flatten layer. A dense layer is stacked before the decision layer.
\end{itemize}

\subsection{Global Performance}

Although the results of the DCASE challenge only report the mean accuracy value, we consider 10 runs to provide not only the mean accuracy value, but also the standard deviation. As it can be seen in Table \ref{tab:results}, all the configurations detailed in Fig.~\ref{fig:squeeze_2d} obtain better accuracy than the DCASE baseline. The contribution of the scSE block is easily justified as \textit{Conv-Residual} gets the lowest performance among the studied configurations. In general, \textit{POST} configurations show a slight improvement compared to the \textit{Standard} configuration. This behaviour differs from what was reported in the original paper, \cite{hu2018squeeze}, in which these blocks were analyzed in the image domain, where the Standard block outperforms the POST block. There is no remarkable difference between \textit{Conv-POST} and \textit{Conv-POST-ELU}. It is also shown that the networks that incorporate the two novel blocks presented in this work, the ones depicted in Figs.~\ref{fig:squeeze_2d}(e) and (f), exhibit the best accuracy values. The shortcut addition at two differente points of the residual block, this is, before and after the scSE block, allowed the network to obtain a more precise classification in this ASC task.

\subsection{Class-wise Performance}

Figure \ref{fig:conf_matrices} shows confusion matrices for each of the analyzed residual blocks in this work. In general, the performance across the different classes is considerably balanced. The ``Public square" class is the one showing the worst performance, tending to be misclassified as ``Street, Pedestrian". Other similar classes such as ``Airport" and ``Shopping mall" or ``Tram" and ``Bus" or ``Metro" tend also to produce common errors in the analyzed networks. 

By analyzing the class-wise performance of the two proposed blocks with respect to the conventional \emph{Conv-Residual} block, substantial improvements are observed. Considering the proposed \emph{Conv-StandardPOST} block, a significant improvement is observed for the classes ``Metro station" and ``Street, Pedestrian". Other classes showing slight improvements are ``Shopping mall", "Park" or "Public square". The class showing the worst relative result was ``Airport".  On the other hand, the second proposed block \emph{Conv-StandardPOST-ELU} provides substantial improvements in ``Street traffic" and ``Park", but other classes like ``Airport" or ``Bus" were degraded.

Finally, when considering the performance of networks implementing SE blocks together, from a general perspective, it is noticed that classes like ``Street, Pedestrian", ``Park" or ``Public square" are improved with respect to the conventional residual network. Only the class ``Airport" shows the best performance in the conventional network, followed by ``Bus". The remaining classes are improved or worsened across all configurations in a degree not as significant as the aforementioned ones.



\begin{figure}[]
    \centering
    \centerline{\includegraphics[scale=0.5]{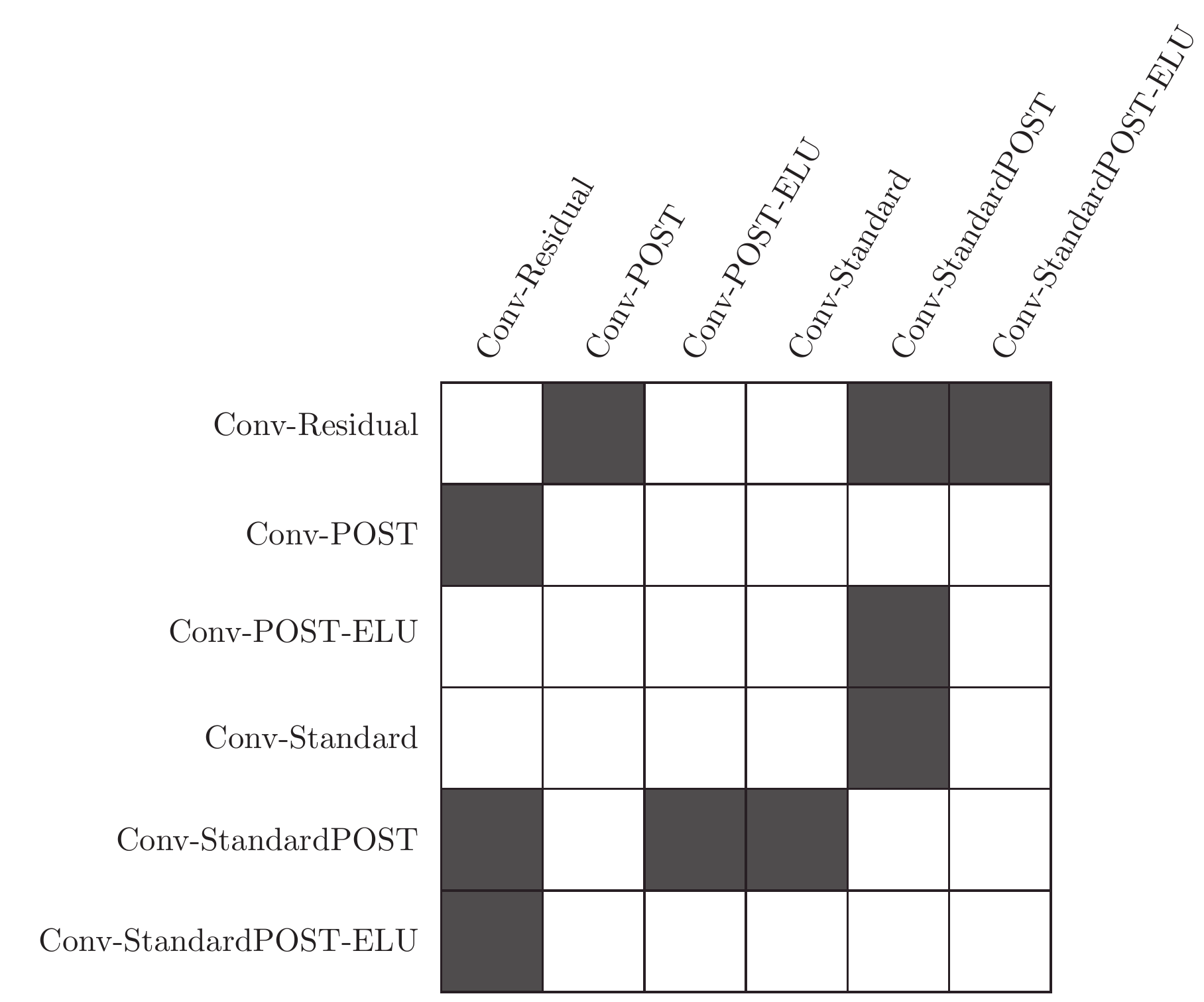}}
    \caption{Pairwise analysis of the studied residual networks using McNemar’s test. Gray cells indicate $p$-values below a 0.05 significance level.}
    \label{fig:test}
\end{figure}

\subsection{Significance Test}

To determine if there are statistically significant differences in the performance of the different blocks analyzed in this work, a McNemar’s test has been carried out \cite{dietterich1998approximate}. This test, which is a paired non-parametric hypothesis test, has been widely recommended for evaluating deep learning models, which are often trained on very large datasets. The test is based on a contingency table created from the results obtained for two methods trained on exactly the same training test and evaluated on the same test set. The null hypothesis of the test is that the performance of the two analyzed systems disagree to the same amount. If the null hypothesis is rejected, there is evidence to suggest that the two systems have different performance when trained on a particular training set. Given a significance level $\alpha$, if $p < \alpha$, there may be sufficient evidence to claim that the two classifiers show different proportions of errors. The result of applying the McNemar’s test to all the available system pairs is shown in 
Fig.~\ref{fig:test}. Gray cells indicate $p$-values below a significance level of 0.05. It is confirmed that the two proposed blocks, \emph{Conv-StandardPOST} and \emph{Conv-StandardPOST-ELU}, show significant differences in performance with respect to all the other blocks but \emph{Conv-POST}, which was the third best performing block. However, no significant differences can be observed between these new blocks, which only differ in the final ELU activation.

\section{Conclusion}\label{sec:conclusion}


The use of squeeze-excitation blocks in convolutional neural networks allows to perform a spatial and channel-wise recalibration of its inner feature maps. This work presented the use of squeeze-excitation residual networks for addressing the acoustic scene classification problem, and presented two novel block configurations that consider residual learning of standard and recalibrated outputs jointly. Results over the well-known DCASE dataset confirm that the proposed blocks provide meaningful improvements by adding a slight architecture modification, outperforming other competing approaches when no data augmentation or model ensembles are considered.

\section*{Acknowledgment}

This project has received funding from the European Union’s Horizon 2020 research and innovation program under grant agreement No 779158. The participation of Javier Naranjo-Alcazar and Dr. Pedro Zuccarello in this work is partially supported by Torres Quevedo fellowships DIN2018-009982 and PTQ-17-09106 respectively from the Spanish Ministry of Science, Innovation and Universities. The participation of Dr. Cobos is supported by FEDER and the Spanish Ministry of Science,
Innovation and Universities under Grant RTI2018-097045-BC21.

\bibliographystyle{unsrt}
\bibliography{references}

\end{document}